\documentclass[twocolumn, twoside]{IEEEtran}

\usepackage{graphicx}
\usepackage{amsmath}
\usepackage{amssymb}
\usepackage{amsthm}

\newtheorem{theorem}{Theorem}
\newtheorem{lemma}{Lemma}
\newtheorem{corollary}{Corollary}
\newtheorem{definition}{Definition}

\begin{document}

\title{The Exponent of a Polarizing Matrix Constructed from the
Kronecker Product}

\author{Myung-Kyu Lee and Kyeongcheol Yang
\thanks{This work was supported by the National Research
Foundation of Korea (NRF) grant funded by the Korea government
(MEST) (No. 2011-0017396).}
\thanks{M.-K. Lee and K. Yang are with the Dept.
   of Electrical Engineering,
   Pohang University of Science and Technology (POSTECH),
   Pohang, Kyungbuk 790-784, Korea (e-mail: {mklee, kcyang}@postech.ac.kr).}
  }
{\markboth{Submitted to IEEE Transactions on Information Theory,
July 27, 2011}{Lee and Yang: The exponent of a polarizing matrix
constructed from the Kronecker product} \maketitle

\begin{abstract}

The asymptotic performance of a polar code under successive
cancellation decoding is determined by the exponent of its
polarizing matrix. We first prove that the partial distances of a
polarizing matrix constructed from the Kronecker product are
simply expressed as a product of those of its component matrices.
We then show that the exponent of the polarizing matrix is shown
to be a weighted sum of the exponents of its component matrices.
These results may be employed in the design of a large polarizing
matrix with high exponent.
\end{abstract}

\begin{keywords}
Polar codes, channel polarization, rate of polarization, partial
distances, exponent, Kronecker product.
\end{keywords}

\section{Introduction}

Channel polarization introduced by Ar{\i}kan \cite{Arikan1} is a
method to construct a class of capacity-achieving codes, called
{\it polar codes}, for symmetric binary-input discrete memoryless
channels (BI-DMCs). Since polar codes are constructed by a
well-defined rule and are provably capacity-achieving, they have
attracted much attention. The probability of block error for polar
coding based on Ar{\i}kan's construction under successive
cancellation (SC) decoding was analyzed by Ar{\i}kan and Telatar
\cite{Arikan2}. Mori and Tanaka employed density evolution in
order to find the frozen bits for polar coding \cite{Mori}.
Recently, Korada {\it et al.} constructed new polar codes using
larger matrices than the $2 \times 2$ matrix proposed by Ar{\i}kan
and analyzed their polarization rate via the partial distances and
exponent \cite{Korada_IT}.

A method to construct polar codes of length $l=l_1 l_2 \cdots l_N$
is to employ a generator matrix of the form $A_{1} \otimes \cdots
\otimes A_{N}$, where $\otimes$ denotes the Kronecker product and
each $A_{i}$ is an $l_i \times l_i$ polarizing matrix
\cite{Korada_ISIT}. One interesting problem is to analyze the
characteristics of such a polarizing matrix. In this paper, we
study the partial distances and the exponent of a polarizing
matrix $A \otimes B$ where $A$ and $B$ are $l_1 \times l_1$ and
$l_2 \times l_2$ polarizing matrices, respectively. We first prove
that the partial distances of $A \otimes B$ are directly
determined by those of $A$ and $B$. We then show that the exponent
of $A \otimes B$ is a weighted sum of the exponents of $A$ and
$B$. These results can be generalized to a polarizing matrix of
the form $A_1 \otimes \cdots \otimes A_N$. Finally, we give design
examples to illustrate that our results may be employed in the
design of a large polarizing matrix with high exponent.

The outline of the paper is as follows. In Section II, we give
some basic notation and definitions, and review briefly the
partial distances and the exponent of a polarizing matrix. In
Section III, we introduce Hamming weight functions associated with
the Kronecker and Hadamard products. Our main results on the
partial distances and the exponent of a polarizing matrix
constructed from the Kronecker product are given in Section IV. In
Section V, some design examples are presented. Finally, we give
some concluding remarks in Section VI.

\section{Preliminaries}

\subsection{Basic Notation and Definitions}

Let ${\mathbb F}$ be a field and ${\mathbb F}^{\,l}$ the
$l$-dimensional vector space of all $l$-tuple vectors over
${\mathbb F}$. Given two vectors
$\bold{a}=[a_{1},a_{2},\ldots,a_{l}]$ and
$\bold{b}=[b_{1},b_{2},\ldots,b_{l}]$, the Haramard product
$\bold{a} \circ \bold{b}$ and the vector addition
$\bold{a}+\bold{b}$ are defined as
\begin{eqnarray*}
\bold{a} \circ \bold{b} &\triangleq& [a_1 b_1, a_2 b_2, \ldots,
a_l b_l],\\
\bold{a} + \bold{b}     &\triangleq& [a_1+b_1, a_2+b_2, \ldots,
a_l+b_l],
\end{eqnarray*}
respectively. Clearly, the vector addition and the Hadamard
product are associative and commutative, that is,
\begin{eqnarray*}
\bold{a} + (\bold{b}+\bold{c}) &=& (\bold{a}+\bold{b})+\bold{c},\\
\bold{a} \circ (\bold{b} \circ \bold{c}) &=& (\bold{a} \circ
\bold{b})
                                                 \circ \bold{c},\\
\bold{a} + \bold{b} &=& \bold{b}+\bold{a},\\
\bold{a} \circ \bold{b} &=& \bold{b} \circ \bold{a}
\end{eqnarray*}
for any $\bold{a}, \bold{b}, \bold{c} \in {\mathbb F}^{\,l}$. It
is also easily checked that the Hadamard product is distributive
over the addition, that is,
\begin{equation*}
\bold{a} \circ (\bold{b}+\bold{c})=
 \bold{a} \circ \bold{b} + \bold{a} \circ \bold{c}
\end{equation*}
for any $\bold{a}, \bold{b}, \bold{c} \in {\mathbb F}^{\,l}$.

For two vectors $\bold{a}=[a_1, a_2, \ldots, a_l]$ and
$\bold{b}=[b_1, b_2, \ldots, b_m]$ over ${\mathbb F}$, the
Kronecker product $\bold{a} \otimes \bold{b}$ is the vector of
length $lm$, given by
\begin{eqnarray*}
\bold{a} \otimes \bold{b} &\triangleq& [a_1 \bold{b}, a_2
\bold{b},\ldots,
a_l \bold{b}]\\
&=&[a_1 b_1, a_1 b_2, \ldots, a_1 b_m,a_2 b_1, a_2 b_2, \ldots,
a_l b_m].
\end{eqnarray*}
The Kronecker product is associative, i.e., $(\bold{a} \otimes
\bold{b}) \otimes \bold{c}=\bold{a} \otimes (\bold{b} \otimes
\bold{c})$ for any $\bold{a} \in {\mathbb F}^{\,l}, \bold{b} \in
{\mathbb F}^{\,m}, \bold{c} \in {\mathbb F}^{\,n}$. It is also
distributive over the addition, that is,
\begin{equation*}
\bold{a} \otimes (\bold{b}+\bold{c})= \bold{a} \otimes
\bold{b}+\bold{a}\otimes \bold{c}
\end{equation*}
for any $\bold{a} \in {\mathbb F}^{\,l}$ and any $\bold{b},
\bold{c} \in {\mathbb F}^{\,m}$.

Given an $m \times n$ matrix $A=(a_{ij})$ and an $r \times s$
matrix $B=(b_{ij})$ over ${\mathbb F}$, the Kronecker product of
$A$ and $B$, denoted by $A \otimes B$, is defined as the $mr
\times ns$ matrix given by
\begin{equation*}
    A \otimes B \triangleq
\left[%
\begin{array}{cccc}
  a_{11}B & a_{12}B & \cdots & a_{1n}B \\
  a_{21}B & a_{22}B & \cdots & a_{2n}B \\
  \vdots  & ~       & \vdots & ~       \\
  a_{m1}B & a_{m2}B & \cdots & a_{mn}B \\
\end{array}%
\right].
\end{equation*}
If we partition $A$ and $B$ on a row basis, that is,
\begin{equation*}
    A =
\left[%
\begin{array}{c}
  \bold{a}_1  \\
  \bold{a}_2  \\
  \vdots   \\
  \bold{a}_m  \\
\end{array}%
\right], B =
\left[%
\begin{array}{c}
  \bold{b}_1  \\
  \bold{b}_2  \\
  \vdots   \\
  \bold{b}_r  \\
\end{array}%
\right]
\end{equation*}
where $\bold{a}_{i}$ and $\bold{b}_{j}$ are the $i$th and $j$th
rows of $A$ and $B$, respectively, then $A \otimes B$ may be
expressed as
\begin{equation*}
    A \otimes B =
\left[%
\begin{array}{c}
  \bold{a}_{1} \otimes \bold{b}_{1}  \\
  \bold{a}_{2} \otimes \bold{b}_{2}  \\
  \vdots                             \\
  \bold{a}_{1} \otimes \bold{b}_{r}  \\
  \bold{a}_{2} \otimes \bold{b}_{1}  \\
  \vdots                             \\
  \bold{a}_{m} \otimes \bold{b}_{r}  \\
\end{array}%
\right].
\end{equation*}
Clearly, the Kronecker product of matrices is associative, that
is,
\begin{equation*}
A \otimes (B \otimes C)=(A \otimes B) \otimes C
\end{equation*}
for any matrices $A, B, C$. For simple notation, let $A^{\otimes
n}$ denote the $n$th Kronecker power of $A$, given by
\begin{equation*}
A^{\otimes n}=\underbrace{A \otimes \cdots \otimes
A}_{n~\text{times}}.
\end{equation*}

\subsection{Partial Distances and Exponent of a Polarizing Matrix}

From now on, we are restricted only to the binary field ${\mathbb
F}_{2} = \{0,1\}$. For a binary vector $\bold{a}$, we denote
$w(\bold{a})$ by its (Hamming) weight, that is, the number of
nonzero components in $\bold{a}$. Let ${\rm supp}(\bold{a})$ be
the support of $\bold{a}=[a_1,a_2,\ldots,a_l]$, given by
\begin{equation*}
    {\rm supp}(\bold{a})=\{1 \leq i \leq l \,|\, a_i \neq 0\}.
\end{equation*}
Clearly, $w(\bold{a})=|{\rm supp}(\bold{a})|$. The (Hamming)
distance $d(\bold{a},\bold{b})$ between two binary vectors
$\bold{a}$ and $\bold{b}$ of length $l$ is defined as the number
of positions at which the corresponding symbols are different in
the two vectors. In particular,
\begin{equation}
d(\bold{a},\bold{b})=w(\bold{a}+\bold{b}). \label{hamming}
\end{equation}

Consider the binary linear code ${\cal C}$ generated by
$\bold{g}_{1},\ldots,\bold{g}_{k} \in {\mathbb F}_{2}^{\,l}$,
denoted by ${\cal C}=\langle \bold{g}_{1},\ldots,\bold{g}_{k}
\rangle$. The minimum distance between ${\cal C}$ and a vector
$\bold{b} \in {\mathbb F}_{2}^{\,l}$, denoted by $d(\bold{b},\cal
C)$, is defined as
\begin{equation*}
d(\bold{b},{\cal C})=\min_{\,\bold{c} \in {\cal
C}}\,d(\bold{b},\bold{c}).
\end{equation*}
The coset of ${\cal C}$ containing $\bold{b}$ is defined as the
set given by
\begin{equation*}
\bold{b}+{\cal C}=\{\bold{b}+\bold{c} \,|\, \bold{c}\in {\cal
C}\}.
\end{equation*}

\medskip
\setcounter{definition}{0}
\begin{definition}[\cite{Korada_IT}]\label{definition1}
Given an $l \times l$ binary matrix
$G=[\bold{g}_{1}^{T},\bold{g}_{2}^{T},\ldots,\bold{g}_{l}^{T}]^{T}$,
the partial distances $D_{G,i}$, $i=1,\ldots,l$ are defined as
\begin{eqnarray*}
    D_{G,i} &\triangleq& d\left(\bold{g}_{i},\langle
                                \bold{g}_{i+1},\ldots,
                                \bold{g}_{l} \rangle
                              \right),~~~i=1,\ldots,l-1\\
    D_{G,l} &\triangleq& d(\bold{g}_l,\bold{0})
\end{eqnarray*}
where $()^{T}$ is the transpose operation and $\bold{0}$ denotes
the all-zero vector.
\end{definition}

\medskip
\setcounter{theorem}{1}

\begin{theorem}[\cite{Korada_IT}]\label{theorem2}
For any BI-DMC and any $l \times l$ polarizing matrix $G$ with
partial distances $\{D_{G,i}\}_{i=1}^{l}$, the rate of
polarization $E(G)$ is given by
\begin{equation*}
    E(G)=\frac{1}{l} \sum_{i=1}^{l} \log_{l}D_{G,i}.
\end{equation*}
\end{theorem}

For convenience, it is referred to as {\it the exponent} of the
matrix $G$ \cite{Korada_IT}. It is known in \cite{Korada_IT} that
when $n$ is sufficiently large, the block error probability of a
polar code constructed by $G^{\otimes n}$ under SC decoding,
$P_e(l^{n})$ can be bounded as
\begin{equation*}
P_e(l^{n}) \leq 2^{-l^{n \beta}}
\end{equation*}
for any positive number $\beta \leq E(G)$. Due to this property,
the exponent of a polarizing matrix can be employed as a
meaningful performance measure of the corresponding polar code
under SC decoding.

\section{Weight Functions Associated with the Kronecker
and Hadamard Products}

The weights of the addition, the Hadamard product and the
Kronecker product of two binary vectors are well-known or easily
computed. The following lemma will be useful in computing the
weight of a more complicated combination of many binary vectors.

\setcounter{lemma}{2}

\begin{lemma} \label{lemma3}
\begin{itemize}
    \item[]
    \item[{\rm i)}] For any $\bold{a}, \bold{b} \in {\mathbb F}_{2}^{\,l}$,
        \begin{equation*}
          w(\bold{a}+\bold{b})= w(\bold{a})+w(\bold{b})
                              -2w(\bold{a} \circ \bold{b}).
        \end{equation*}
   \item[{\rm ii)}] For any $\bold{a} \in {\mathbb F}_{2}^{\,l}$, $\bold{b} \in
                    {\mathbb F}_{2}^{\,m}$,
        \begin{equation*}
          w(\bold{a}\otimes\bold{b})= w(\bold{a})w(\bold{b}).
        \end{equation*}
   \item[{\rm iii)}] For any $\bold{a}, \bold{b} \in
                     {\mathbb F}_{2}^{\,l}$,
        \begin{equation*}
             w(\bold{a}\circ\bold{b}) \leq \min(w(\bold{a}),w(\bold{b}))
        \end{equation*}
        with equality iff ${\rm supp}(\bold{a}) \subset {\rm supp}(\bold{b})$
        or vice versa.
   \item[{\rm iv)}] For any
                    $\bold{a}_1, \bold{a}_2 \in {\mathbb F}_{2}^{\,l}$ and any
                    $\bold{b}_1, \bold{b}_2 \in {\mathbb F}_{2}^{\,m}$,
        \begin{equation*}
        w((\bold{a}_1 \otimes \bold{b}_1) \circ (\bold{a}_2 \otimes \bold{b}_2))
        =w(\bold{a}_1 \circ \bold{a}_2)
         w(\bold{b}_1 \circ \bold{b}_2).
        \end{equation*}
    \item[{\rm v)}] For any
                    $\bold{a}_1, \bold{a}_2 \in {\mathbb F}_{2}^{\,l}$ and any
                    $\bold{b}_1, \bold{b}_2 \in {\mathbb F}_{2}^{\,m}$,
        \begin{eqnarray*}
        w(\bold{a}_1 \otimes \bold{b}_1+\bold{a}_2 \otimes \bold{b}_2)
        &=&w(\bold{a}_1) w(\bold{b}_1)+w(\bold{a}_2)
        w(\bold{b}_2)\\
        &&-2w(\bold{a}_1 \circ \bold{a}_2)w(\bold{b}_1 \circ \bold{b}_2).
        \end{eqnarray*}
\end{itemize}
\end{lemma}
\begin{IEEEproof}
i), ii) and iii) are obvious. iv) comes from ii) and the fact
\begin{equation*}
    (\bold{a}_1 \otimes \bold{b}_1) \circ
    (\bold{a}_2 \otimes \bold{b}_2) =
    (\bold{a}_1 \circ \bold{a}_2) \otimes
    (\bold{b}_1 \circ \bold{b}_2).
\end{equation*}
v) is directly obtained by applying i) and iv).
\end{IEEEproof}

\bigskip

The following three lemmas can be easily derived by applying the
mathematical induction and Lemma \ref{lemma3}.

\medskip

\begin{lemma}\label{lemma4}
For any $\bold{a}_{1},\ldots,\bold{a}_{K} \in {\mathbb
F}_{2}^{\,l}$,
\begin{eqnarray*}
&&w( \bold{a}_{1}+\cdots+\bold{a}_{K} )
=\sum_{i=1}^{K}w(\bold{a}_i)\\
&&~~~~~~~~~~~~-2\cdot\sum_{1\leq i_1 < i_2 \leq K}
       w(\bold{a}_{i_1}\circ \bold{a}_{i_2})\\
&&~~~~~~~~~~~~+\,4\cdot\sum_{1\leq i_1 < i_2 <i_3 \leq K}
       w(\bold{a}_{i_1}\circ \bold{a}_{i_2} \circ \bold{a}_{i_3})\\
&&~~~~~~~~~~~~+\cdots+(-2)^{K-1}
   w(\bold{a}_1 \circ \bold{a}_2 \circ \cdots \circ \bold{a}_K).
\end{eqnarray*}
\end{lemma}

\medskip

\begin{lemma}\label{lemma5}
For any $\bold{a}_1,\ldots,\bold{a}_K \in {\mathbb F}_{2}^{\,l}$
and any $\bold{b}_1,\ldots,\bold{b}_K \in {\mathbb F}_{2}^{\,m}$,
\begin{equation*}
    w( (\bold{a}_1 \otimes \bold{b}_1) \circ \cdots \circ
       (\bold{a}_K \otimes \bold{b}_K ) )
    =w(\bold{a}_1 \circ \cdots \circ \bold{a}_K)
     w(\bold{b}_1 \circ \cdots \circ \bold{b}_K).
\end{equation*}
\end{lemma}

\medskip

\begin{lemma}\label{lemma6}
For any $\bold{a}_1,\ldots,\bold{a}_K \in {\mathbb F}_{2}^{\,l}$
and any $\bold{b}_1,\ldots,\bold{b}_K \in {\mathbb F}_{2}^{\,m}$,
\begin{eqnarray*}
&&w\left(\sum_{i=1}^{K} \bold{a}_{i} \otimes \bold{b}_{i} \right)
 =   \sum_{i=1}^{K}w(\bold{a}_i)w(\bold{b}_i)\\
&&~-2\cdot\sum_{1\leq i_1 < i_2 \leq K}
     w(\bold{a}_{i_1}\circ \bold{a}_{i_2})
     w(\bold{b}_{i_1}\circ \bold{b}_{i_2})\\
&&~+\,4\cdot\sum_{1\leq i_1 < i_2 <i_3 \leq K}
     w(\bold{a}_{i_1}\circ \bold{a}_{i_2} \circ \bold{a}_{i_3})
     w(\bold{b}_{i_1}\circ \bold{b}_{i_2} \circ \bold{b}_{i_3})\\
&&~+ \cdots+(-2)^{K-1}
   w(\bold{a}_1 \circ \cdots \circ \bold{a}_K)
   w(\bold{b}_1 \circ \cdots \circ \bold{b}_K).
\end{eqnarray*}
\end{lemma}

\medskip

In order to analyze the partial distances of a polarizing matrix
$A \otimes B$ in the next section, we need to introduce two kinds
of weight functions, that is, the {\it weight exclusion function}
and the {\it weight difference function}. More specifically, these
two functions will be employed in proving that the partial
distances of $A \otimes B$ are expressed as a product of those of
$A$ and $B$.

\setcounter{definition}{6}

\begin{definition} \label{definition7}
Let $\bold{a}_{i} \in {\mathbb F}_{2}^{\,l}$ for $1 \leq i \leq
K$. For $K=1$, let $f_{1}(\bold{a}_{1})=w(\bold{a}_{1})$. For $K
\geq 2$, the {\it weight exclusion function}
$f_{K}(\bold{a}_1;\bold{a}_2,$ $\ldots,\bold{a}_K)$ is defined as
\begin{eqnarray*}
&&f_{K}(\bold{a}_1;\bold{a}_2,\ldots,\bold{a}_K) \triangleq
w(\bold{a}_1)
   -\sum_{i=2}^{K}w(\bold{a}_{1} \circ \bold{a}_{i})\\
&&~~~~~~~~~+\sum_{2 \leq i_1 < i_2 \leq K}
    w(\bold{a}_{1} \circ \bold{a}_{i_1} \circ \bold{a}_{i_2})\\
&&~~~~~~~~~- \sum_{2 \leq i_1 < i_2 <i_3 \leq K}
    w(\bold{a}_{1} \circ \bold{a}_{i_1} \circ \bold{a}_{i_2}
                   \circ \bold{a}_{i_3})\\
&&~~~~~~~~~+\cdots+(-1)^{K-1}
    w(\bold{a}_1 \circ \bold{a}_2 \circ \cdots \circ
          \bold{a}_K).
\end{eqnarray*}
\end{definition}

\medskip
\setcounter{lemma}{7}

\begin{lemma}\label{lemma8}
For any $\bold{a}_1,\ldots,\bold{a}_K \in {\mathbb F}_{2}^{\,l}$,
\begin{eqnarray*}
   f_{K}(\bold{a}_1;\bold{a}_2,\ldots,\bold{a}_K)
    =w(\bold{a}_{1}       \circ
       \bar{\bold{a}}_{2} \circ \cdots \circ
       \bar{\bold{a}}_{K})
\end{eqnarray*}
where $\bar{\bold{a}}$ denotes the complement of
$\bold{a}=[a_1,a_2,\ldots,a_l]$, that is,
\begin{equation*}
\bar{\bold{a}}=[1+a_1,1+a_2,\ldots,1+a_l].
\end{equation*}
In particular, $f_{K}(\bold{a}_1;\bold{a}_2,\ldots,\bold{a}_K)
\geq 0$.
\end{lemma}
\begin{IEEEproof}
Let $S_{i}$ be the support of $\bold{a}_{i}$. Clearly,
$w(\bar{\bold{a}}_{i})=l-|S_i|=\left|S_{i}^{\,\rm C}\right|$,
where $S^{\,\rm C}$ denotes the complement set of $S$. Using the
inclusion-exclusion principle \cite{IE_principle}, we have
\begin{eqnarray*}
 &&  f_{K}(\bold{a}_1;\bold{a}_2,\ldots,\bold{a}_K)\\
 &&~~~=|S_1|-|(S_1 \cap S_2) \cup (S_1 \cap S_3) \cup \cdots
                           \cup (S_1 \cap S_K) | \\
 &&~~~=|S_1|-|S_1 \cap (S_2 \cup S_3 \cup \cdots \cup S_K)|\\
 &&~~~=|S_1 \cap (S_2 \cup \cdots \cup S_K)^{\rm C}|\\
 &&~~~=|S_1 \cap S_{2}^{\,\rm C} \cap \cdots \cap S_{K}^{\,\rm
C}|\\
 &&~~~=w(\bold{a}_{1}       \circ
         \bar{\bold{a}}_{2} \circ \cdots \circ
         \bar{\bold{a}}_{K}).
\end{eqnarray*}
\end{IEEEproof}

\medskip

\begin{lemma}\label{lemma9}
For any $\bold{a}_1,\ldots,\bold{a}_K \in {\mathbb F}_{2}^{\,l}$,
\begin{eqnarray}
 f_{1}(\bold{a}_1) &=&
   f_{K}(\bold{a}_1; \bold{a}_2,\bold{a}_3,\ldots,\bold{a}_K)
         \nonumber \\
 &&+\,f_{K-1}(\bold{a}_1 \circ \bold{a}_2;\bold{a}_3,\ldots,
              \bold{a}_K) \nonumber \\
 &&+\cdots+f_{K-1}(\bold{a}_1 \circ \bold{a}_K;\bold{a}_2,
                   \ldots,          \bold{a}_{K-1})
   \nonumber \\
 &&+\,f_{K-2}(\bold{a}_1 \circ \bold{a}_2 \circ \bold{a}_3;
             \bold{a}_4,\ldots,\bold{a}_K) \nonumber \\
 &&+\cdots+f_{K-2}(\bold{a}_1 \circ \bold{a}_{K-1} \circ \bold{a}_{K};
                   \bold{a}_2,\ldots,\bold{a}_{K-2})  \nonumber \\
 &&+\,\cdots+f_{1}  (\bold{a}_1 \circ \bold{a}_2 \circ \bold{a}_3 \circ
                 \cdots \circ \bold{a}_{K}).\label{functionf1}
\end{eqnarray}
\end{lemma}
\begin{IEEEproof}
Note that $\bold{a}=\bold{a} \circ (\bold{b}+\bar{\bold{b}})$ for
any $\bold{a},\bold{b} \in {\mathbb F}_{2}^{\,l}$, since
$\bold{b}+\bar{\bold{b}}=[1,1,\ldots,1]$. Therefore,
\begin{eqnarray*}
w(\bold{a})&=&w(\bold{a} \circ \bold{b})+
              w(\bold{a} \circ \bar{\bold{b}})\\
&=&\sum_{\bold{x} \in \{\bold{b},\bar{\bold{b}}\}}
w(\bold{a}\circ\bold{x}).
\end{eqnarray*}
Applying the above relation to $\bold{a}_1$ repeatedly, we have
\begin{equation*}
w(\bold{a}_{1})= \sum_{\bold{x}_2 \in
\{\bold{a}_2,\bar{\bold{a}_2}\}} \sum_{\bold{x}_3 \in
\{\bold{a}_3,\bar{\bold{a}_3}\}} \cdots \sum_{\bold{x}_K \in
\{\bold{a}_K,\bar{\bold{a}_K}\}} w(\bold{a}_{1} \circ \bold{x}_2
\circ \cdots \circ \bold{x}_{K}).
\end{equation*}
Using the commutativity of the Hadamard product and the definition
of $f_{K}$, we complete the proof.
\end{IEEEproof}

\bigskip
\setcounter{definition}{9}

\begin{definition}\label{definition10}
Let $\bold{a}_1,\ldots,\bold{a}_K \in {\mathbb F}_{2}^{\,l}$ and
$\bold{b}_1,\ldots,\bold{b}_K \in {\mathbb F}_{2}^{\,m}$. For
$K=1$, let $g_{1}(\bold{a}_1;\bold{b}_1)=0$. For $K \geq 2$, the
{\it weight difference function} is defined as
\begin{eqnarray}
 &&g_{K}(\bold{a}_{1};\bold{a}_{2},\ldots,\bold{a}_{K};
   \bold{b}_{1};\bold{b}_{2},\ldots,\bold{b}_{K}) \nonumber \\
 &&~~~~~~~~~~\triangleq
  w\left( \sum_{i=1}^{K} \bold{a}_{i} \otimes \bold{b}_{i} \right)
 -w(\bold{a}_1 \otimes \bold{b}_1).
 \label{function_M1}
\end{eqnarray}
\end{definition}

\medskip

Note that $g_{K}$ can be expressed as a linear combination of
$f_{i}$'s. For example, if we take $K=2$, we get
\begin{eqnarray}
&&g_{2}(\bold{a}_{1};\bold{a}_{2};\bold{b}_{1};\bold{b}_2) \nonumber \\
&&=w(\bold{a}_2)w(\bold{b}_2)-2w(\bold{a}_{1}\circ\bold{a}_{2})
w(\bold{b}_{1}\circ\bold{b}_{2}) \nonumber \\
&&=w(\bold{a}_2)w(\bold{b}_2)+
[w(\bold{a}_{1}+\bold{a}_{2})-w(\bold{a}_1)-w(\bold{a}_2)]
w(\bold{b}_{1}\circ\bold{b}_{2}) \nonumber \\
&&=[w(\bold{a}_1+\bold{a}_2)-w(\bold{a}_1)]
    w(\bold{b}_1\circ\bold{b}_2)\nonumber \\
&&~~~+w(\bold{a}_2) [w(\bold{b}_2)-w(\bold{b}_1\circ\bold{b}_2)]
\nonumber \\
&&=[w(\bold{a}_1+\bold{a}_2)-w(\bold{a}_1)]
f_1(\bold{b}_1\circ\bold{b}_2)
+w(\bold{a}_2)f_{2}(\bold{b}_{2};\bold{b}_{1}). \label{g2}
\end{eqnarray}
Such an expression as in (\ref{g2}) plays a key role in proving
that $g_K \geq 0$ under some conditions.

\medskip
\setcounter{lemma}{10}

\begin{lemma}\label{lemma11}
For a positive integer $K$, let $\bold{a}_1,\ldots,\bold{a}_{K}
\in {\mathbb F}_{2}^{\,l}$ such that $w(\bold{a}_1) \leq
w(\bold{a}_1+\sum_{i=2}^{K} \epsilon_{i}\bold{a}_{i})$ for any
$\epsilon_{i} \in {\mathbb F}_2$. Then
\begin{equation*}
    g_{K}(\bold{a}_1;\bold{a}_2,\ldots,\bold{a}_K;
          \bold{b}_1;\bold{b}_2,\ldots,\bold{b}_K) \geq 0
\end{equation*}
for any $\bold{b}_1,\bold{b}_2,\ldots,\bold{b}_{K} \in {\mathbb
F}_{2}^{\,m}$. In particular, $g_{K}=0$ if $\bold{b}_{i}=\bold{0}$
for all $i \geq 2$.
\end{lemma}
\begin{IEEEproof}
We first show that $g_K$ can be expressed as a linear combination
of $f_{i}$'s. It is true for $g_1$ by definition. The expression
for $g_2$ is given in (\ref{g2}). In order to illustrate such an
expression by a more example, if we take $K=3$, we have
\begin{eqnarray}
&&g_{3}(\bold{a}_1;\bold{a}_2,\bold{a}_3;\bold{b}_{1};\bold{b}_2,\bold{b}_3)
\nonumber \\
&&= w(\bold{a}_2)w(\bold{b}_2)
   +w(\bold{a}_3)w(\bold{b}_3)
   -2w(\bold{a}_1\circ\bold{a}_2)w(\bold{b}_1\circ\bold{b}_2)
   \nonumber\\
&&~~-\,2w(\bold{a}_1\circ\bold{a}_3)w(\bold{b}_1\circ\bold{b}_3)
     -2w(\bold{a}_2\circ\bold{a}_3)w(\bold{b}_2\circ\bold{b}_3)
    \nonumber\\
&&~~+\,4w(\bold{a}_1\circ\bold{a}_2\circ\bold{a}_3)
       w(\bold{b}_1\circ\bold{b}_2\circ\bold{b}_3).\label{g3}
\end{eqnarray}
By Lemma \ref{lemma4}, we get
\begin{eqnarray}
&&4w(\bold{a}_1 \circ \bold{a}_2 \circ
\bold{a}_3) \nonumber \\
&&~~~=w(\bold{a}_1+\bold{a}_2+\bold{a}_3)
-w(\bold{a}_1)-w(\bold{a}_2)-w(\bold{a}_3) \nonumber \\
&&~~~~~+2w(\bold{a}_1 \circ \bold{a}_2)+2w(\bold{a}_2 \circ
\bold{a}_3)+2w(\bold{a}_1 \circ \bold{a}_3). \label{4w}
\end{eqnarray}
Plugging (\ref{4w}) into (\ref{g3}), we obtain
\begin{eqnarray}
&&g_{3}(\bold{a}_1;\bold{a}_2,\bold{a}_3;\bold{b}_{1};\bold{b}_2,\bold{b}_3)
~~~~~~~~~~~~~~~~~~~~~~~~~~~~~~~~~~~~~~~`
\nonumber \\
&&~~=
   w(\bold{a}_2)
   [w(\bold{b}_2)-w(\bold{b}_1\circ\bold{b}_2\circ\bold{b}_3)]
   \nonumber\\
&&~~~~~+\,w(\bold{a}_3)
   [w(\bold{b}_3)-w(\bold{b}_1\circ\bold{b}_2\circ\bold{b}_3)]
   \nonumber\\
&&~~~~~-\,2w(\bold{a}_1\circ\bold{a}_2)
   [w(\bold{b}_1\circ\bold{b}_2)
    -w(\bold{b}_1\circ\bold{b}_2\circ\bold{b}_3)]
   \nonumber\\
&&~~~~~-\,2w(\bold{a}_1\circ\bold{a}_3)
   [w(\bold{b}_1\circ\bold{b}_3)
    -w(\bold{b}_1\circ\bold{b}_2\circ\bold{b}_3)]
   \nonumber\\
&&~~~~~-\,2w(\bold{a}_2\circ\bold{a}_3)
   [w(\bold{b}_2\circ\bold{b}_3)
    -w(\bold{b}_1\circ\bold{b}_2\circ\bold{b}_3)]
   \nonumber\\
&&~~~~~+\,[w(\bold{a}_1+\bold{a}_2+\bold{a}_3)-w(\bold{a}_1)]
      w(\bold{b}_1\circ\bold{b}_2\circ\bold{b}_3)\nonumber.
\end{eqnarray}
From the definition of $f_K$, we have
\begin{eqnarray*}
f_1(\bold{b}_1 \circ \bold{b}_2 \circ \bold{b}_3)&=&w(\bold{b}_1
\circ \bold{b}_2 \circ \bold{b}_3) \\
f_{2}(\bold{b}_1 \circ
\bold{b}_2;\bold{b}_3)&=&w(\bold{b}_1\circ\bold{b}_2)
-w(\bold{b}_1\circ\bold{b}_2\circ\bold{b}_3)\\
f_{2}(\bold{b}_1 \circ
\bold{b}_3;\bold{b}_2)&=&w(\bold{b}_1\circ\bold{b}_3)
-w(\bold{b}_1\circ\bold{b}_2\circ\bold{b}_3)\\
f_{2}(\bold{b}_2 \circ
\bold{b}_3;\bold{b}_1)&=&w(\bold{b}_2\circ\bold{b}_3)
-w(\bold{b}_1\circ\bold{b}_2\circ\bold{b}_3)\\
f_{3}(\bold{b}_2;
\bold{b}_1,\bold{b}_3)&=&w(\bold{b}_2)-w(\bold{b}_1\circ\bold{b}_2)
-w(\bold{b}_2\circ\bold{b}_3) \nonumber \\
&&+\,w(\bold{b}_1\circ\bold{b}_2\circ\bold{b}_3)\\
f_{3}(\bold{b}_3;
\bold{b}_1,\bold{b}_2)&=&w(\bold{b}_3)-w(\bold{b}_1\circ\bold{b}_3)
-w(\bold{b}_2\circ\bold{b}_3)\nonumber \\
&&+\,w(\bold{b}_1\circ\bold{b}_2\circ\bold{b}_3).
\end{eqnarray*}
Using these relations and the relation $-2w(\bold{a}_i \circ
\bold{a}_j)=w(\bold{a}_i+\bold{a}_j)-w(\bold{a}_i)-w(\bold{a}_j)$,
we get
\begin{eqnarray*}
&&g_{3}(\bold{a}_1;\bold{a}_2,\bold{a}_3;\bold{b}_{1};\bold{b}_2,\bold{b}_3)
\nonumber \\
&&~~~= [w(\bold{a}_1+\bold{a}_2+\bold{a}_3)-w(\bold{a}_1)]
    f_{1}(\bold{b}_1\circ\bold{b}_2\circ\bold{b}_3)\\
&&~~~~~~+\,[w(\bold{a}_1+\bold{a}_2)-w(\bold{a}_1)]
    f_{2}(\bold{b}_1 \circ \bold{b}_2;\bold{b}_3)  \\
&&~~~~~~+\,[w(\bold{a}_1+\bold{a}_3)-w(\bold{a}_1)]
    f_{2}(\bold{b}_1 \circ \bold{b}_3;\bold{b}_2)  \\
&&~~~~~~+\,w(\bold{a}_2+\bold{a}_3)
    f_{2}(\bold{b}_2 \circ \bold{b}_3;\bold{b}_1)  \\
&&~~~~~~+\,w(\bold{a}_2)f_{3}(\bold{b}_2;\bold{b}_1,\bold{b}_3)
    +w(\bold{a}_3)f_{3}(\bold{b}_3;\bold{b}_1,\bold{b}_2).  \\
\end{eqnarray*}
In the same procedure as above, it is possible to express $g_K$ as
\begin{eqnarray}
 &&g_{K}(\bold{a}_{1};\bold{a}_{2},\ldots,\bold{a}_{K};
    \bold{b}_{1};\bold{b}_{2},\ldots,\bold{b}_{K})
   \nonumber \\
&&~~=[ w(\bold{a}_1+\bold{a}_2+\cdots+\bold{a}_K)
       -w(\bold{a}_1)] \nonumber \\
&&~~~~~~\cdot
    \,f_{1}(\bold{b}_1 \circ \bold{b}_2 \circ \cdots
                     \circ \bold{b}_{K})
    \nonumber\\
&&~~~~~+[ w(\bold{a}_1+\bold{a}_2+\cdots+\bold{a}_{K-1})
        -w(\bold{a}_1)] \nonumber \\
&&~~~~~~~~\cdot \,f_{2}(\bold{b}_1 \circ \bold{b}_2 \circ \cdots
\circ
    \bold{b}_{K-1};\bold{b}_{K})\nonumber\\
&&~~~~~+[w(\bold{a}_1+\cdots+\bold{a}_{K-2}+\bold{a}_{K})
       -w(\bold{a}_1)]\nonumber \\
&&~~~~~~~~\cdot \,f_{2}(\bold{b}_1 \circ \cdots \circ
\bold{b}_{K-2}
                   \circ \bold{b}_{K};\bold{b}_{K-1}) \nonumber\\
&&~~~~~+\cdots+w(\bold{a}_2+\bold{a}_3+\cdots+\bold{a}_{K})
\nonumber \\
&&~~~~~~~~~~~~~~~\cdot  \,f_{2}(\bold{b}_{2} \circ \bold{b}_{3}
\circ \cdots\ \circ
        \bold{b}_{K};\bold{b}_{1})\nonumber\\
&&~~~~~+[w(\bold{a}_1+\bold{a}_2+\cdots+\bold{a}_{K-2})
       -w(\bold{a}_1)] \nonumber \\
&&~~~~~~~~~\cdot  \,f_{3}(\bold{b}_1 \circ \bold{b}_2 \circ \cdots
\circ
        \bold{b}_{K-2};  \bold{b}_{K-1},\bold{b}_{K})\nonumber\\
&&~~~~~+\cdots+w(\bold{a}_3+\bold{a}_4+\cdots+\bold{a}_{K})
\nonumber \\
&&~~~~~~~~~~~~~~~\cdot  \,f_{3}(\bold{b}_3 \circ \bold{b}_4 \circ
\cdots \circ
        \bold{b}_K;\bold{b}_1,\bold{b}_2)\nonumber\\
&&~~~~~+\cdots+w(\bold{a}_2)
  \,f_{K}(\bold{b}_2; \bold{b}_1,\bold{b}_3,\ldots,\bold{b}_K)\nonumber \\
&&~~~~~+\cdots
   +w(\bold{a}_K)
  \,f_{K}(\bold{b}_K; \bold{b}_1,\bold{b}_2,\ldots,\bold{b}_{K-1}).
  \label{gk}
\end{eqnarray}
As a second step, we note that the first factor in each term of
$g_K$ is larger than or equal to $0$ by the assumption on
$\bold{a}_1,\ldots,\bold{a}_K$ and $f_i \geq 0$ for any $1 \leq i
\leq K$ by Lemma \ref{lemma8}. Therefore, we complete the proof.
\end{IEEEproof}

\section{Main Results}

Let $A$ be an $l_1 \times l_1$ polarizing matrix with partial
distances $\{D_{A,i}\}_{i=1}^{l_1}$ and $B$ an $l_2 \times l_2$
polarizing matrix with partial distances
$\{D_{B,i}\}_{i=1}^{l_2}$, given by
\begin{equation*}
A=\left[%
\begin{array}{c}
  \bold{a}_1 \\
  \bold{a}_2 \\
  \vdots \\
  \bold{a}_{l_1} \\
\end{array}%
\right],~~~~~
B=\left[%
\begin{array}{c}
  \bold{b}_1 \\
  \bold{b}_2 \\
  \vdots \\
  \bold{b}_{l_2} \\
\end{array}%
\right]
\end{equation*}
where $\bold{a}_{i}$ is the $i$th row of $A$ and $\bold{b}_{j}$ is
the $j$th row of $B$. Note that $A \otimes B$ is an $l_1 l_2
\times l_1 l_2$ polarizing matrix and every integer $k$ with $1
\leq k \leq l_1 l_2$ can be uniquely expressed as $k=(i-1)l_2 + j$
with $1 \leq i \leq l_1$ and $1 \leq j \leq l_2$. Our first
problem is to determine the partial distances of the polarizing
matrix $A \otimes B$ in terms of those of $A$ and $B$.

\medskip

\setcounter{theorem}{11}

\begin{theorem}\label{theorem12}
The partial distances of the polarizing matrix $A \otimes B$ are
given by
\begin{eqnarray*}
D_{A \otimes B,\,(i-1)l_2+j}=D_{A,i} \cdot D_{B,j}
\end{eqnarray*}
for $1 \leq i \leq l_1$ and $1 \leq j \leq l_2$.
\end{theorem}
\begin{IEEEproof}
We divide our problem into two cases depending on the index $i$.

Case 1) $i=1$: Using the relation in (\ref{hamming}), the $j$th
partial distance of $A \otimes B$ is given by
\begin{eqnarray}
D_{A \otimes B,\,j} =
\min_{\bold{x}_{1},\tilde{\bold{x}}_{2},\ldots,
                   \tilde{\bold{x}}_{l_1}}
w\left(\bold{a}_{1} \otimes \bold{x}_{1} + \sum_{k=2}^{l_1}
       \bold{a}_{k} \otimes \tilde{\bold{x}}_{k} \right),\nonumber \\
1 \leq j \leq l_2 \label{pd1}
\end{eqnarray}
where $\bold{x}_{1} \in \bold{b}_{j}+\langle
\bold{b}_{j+1},\bold{b}_{j+2},\ldots,\bold{b}_{l_2} \rangle$, and
$\tilde{\bold{x}}_{2}, \tilde{\bold{x}}_{3},\ldots,
\tilde{\bold{x}}_{l_1} \in \langle
\bold{b}_{1},\bold{b}_{2},\ldots,\bold{b}_{l_2} \rangle$. Let
$\bold{a}_{\rm D} \in \bold{a}_{1}+\langle
\bold{a}_{2},\bold{a}_{3},\ldots,\bold{a}_{l_1} \rangle$ be a
binary vector with minimum weight $D_{A,1}$, i.e.,
$D_{A,1}=w(\bold{a}_{\rm D})$. Then
\begin{equation*}
\bold{a}_{\rm D}=\bold{a}_{1}+\sum_{k=2}^{l_1}
\epsilon_{k}\bold{a}_{k}
\end{equation*}
with $\epsilon_{k} \in {\mathbb F}_{2}$ for $2 \leq k \leq l_1$
and the partial distance $D_{A \otimes B, j}$ in (\ref{pd1}) may
be rewritten as
\begin{eqnarray}
&&D_{A \otimes B,\,j} \nonumber \\
&&= \min_{\bold{x}_{1},\bold{x}_{2},\ldots,
                              \bold{x}_{l_1}}
w\left(\bold{a}_{\rm D} \otimes \bold{x}_{1} + \sum_{k=2}^{l_1}
       \bold{a}_{k} \otimes \bold{x}_{k} \right)
\label{pd2} \\
&&=\min_{\bold{x}_1} \left (\min_{\bold{x}_2,\ldots,
                            \bold{x}_{l_1}}
w\left(\bold{a}_{\rm D} \otimes \bold{x}_{1} + \sum_{k=2}^{l_1}
       \bold{a}_{k} \otimes \bold{x}_{k} \right) \right)~~~~
\label{pd3}
\end{eqnarray}
where $\bold{x}_{k}=\tilde{\bold{x}}_{k}+\epsilon_{k}\bold{x}_{1}$
for $k=2,\ldots,l_1$. Using the weight difference function $g_{K}$
in Definition \ref{definition10}, we may express $w(\cdot)$ in
(\ref{pd3}) as follows:
\begin{eqnarray*}
&&w\left(\bold{a}_{\rm D} \otimes \bold{x}_1
  +\sum_{k=2}^{l_1}\bold{a}_{k} \otimes \bold{x}_{k} \right)
  \nonumber\\
&&=w(\bold{a}_{\rm D} \otimes \bold{x}_1)
  +g_{l_1}(\bold{a}_{\rm D};\bold{a}_{2},\ldots,\bold{a}_{l_1};
         \bold{x}_1      ;\bold{x}_{2},\ldots,\bold{x}_{l_1}).
\end{eqnarray*}
By the choice of $\bold{a}_{\rm D}$ with  $w(\bold{a}_{\rm
D})=D_{A,1}$ and Lemma \ref{lemma11}, it is easily checked that
for any $\bold{x}_1 \in \bold{b}_{j}+\langle
\bold{b}_{j+1},\ldots,\bold{b}_{l_2} \rangle$ and any
$\bold{x}_{2},\ldots,\bold{x}_{l_1} \in \langle
\bold{b}_{1},\ldots,\bold{b}_{l_2} \rangle$,
\begin{eqnarray*}
 g_{l_1}(\bold{a}_{\rm D};\bold{a}_{2},\ldots,\bold{a}_{l_1};
         \bold{x}_1;\bold{x}_{2},\ldots,\bold{x}_{l_1}) \geq 0
\end{eqnarray*}
where the equality holds if $\bold{x}_k=\bold{0}$ for all $k \geq
2$. Therefore, for a given binary vector $\bold{x}_1$
\begin{eqnarray*}
\min_{\bold{x}_2,\ldots,\bold{x}_{l_1}} w\left(\bold{a}_{\rm D}
\otimes \bold{x}_{1} + \sum_{k=2}^{l_1}
           \bold{a}_{k} \otimes \bold{x}_{k} \right)
&=& w(\bold{a}_{\rm D} \otimes \bold{x}_1) \nonumber \\
&=& w(\bold{a}_{\rm D})w(\bold{x}_1) \nonumber \\
&=& D_{A,1}\cdot w(\bold{x}_1).
\end{eqnarray*}
This relation reduces (\ref{pd3}) to
\begin{eqnarray*}
D_{A \otimes B,\,j} &=& \min_{\bold{x}_1}~D_{A,1}\cdot
w(\bold{x}_1) \\
&=& D_{A,1} \cdot \min_{\bold{x} \in \langle
\bold{b}_{j+1},\ldots, \bold{b}_{l_2} \rangle}w(\bold{b}_j+\bold{x})\\
&=&D_{A,1} \cdot D_{B,j}
\end{eqnarray*}
for any $1 \leq j \leq l_2$.

\smallskip

Case 2) $i \geq 2$: Let $A^{(i)}$ be the $(l_1-i+1)\times l_1$
submatrix of $A$, given by
\[A^{(i)}=[\bold{a}_{i}^{T},\bold{a}_{i+1}^{T},\ldots,
          \bold{a}_{l_1}^{T}]^{T}.\]
Then $D_{A \otimes B,(i-1)l_2+j}=D_{A^{(i)} \otimes B,j}$. In a
similar approach as in Case 1), we have
\begin{equation*}
D_{A^{(i)} \otimes B, j}=D_{A^{(i)},1} \cdot D_{B,j}
\end{equation*}
for any $1 \leq j \leq l_2$. Note that the first factor
$D_{A^{(i)},1}$ is exactly equal to $D_{A,i}$. Therefore, we
complete the proof.
\end{IEEEproof}

\bigskip

\begin{theorem} \label{theorem13}
The exponent of the polarizing matrix $A \otimes B$ is given by
\begin{eqnarray*}
E(A \otimes B)= \frac{E(A)}{\log_{l_1}l_1l_2}+
\frac{E(B)}{\log_{l_2}l_1l_2}.
\end{eqnarray*}
\end{theorem}
\begin{IEEEproof}
By Theorems \ref{theorem2} and \ref{theorem12}, we have
\begin{eqnarray*}
E(A \otimes B)&=&\frac{1}{l_1 l_2}
\sum_{i=1}^{l_1}\sum_{j=1}^{l_2} \log_{l_1 l_2}
D_{A\otimes B,(i-1)l_2+j}\\
&=&\frac{1}{l_1 l_2} \sum_{i=1}^{l_1}\sum_{j=1}^{l_2} \log_{l_1
l_2} D_{A,i}\cdot D_{B,j}\\
              &=& \frac{1}{l_1 l_2}\sum_{i=1}^{l_1}\sum_{j=1}^{l_2}
(\log_{l_1 l_2} D_{A,i}+\log_{l_1 l_2} D_{B,j})\\
&=& \frac{1}{l_1}\sum_{i=1}^{l_1} \log_{l_1 l_2} D_{A,i}+
    \frac{1}{l_2}\sum_{j=1}^{l_2} \log_{l_1 l_2} D_{B,j}\\
&=& \frac{1}{l_1}\sum_{i=1}^{l_1}
                 \frac{\log_{l_1}D_{A,i}}{\log_{l_1}l_1l_2}+
    \frac{1}{l_2}\sum_{j=1}^{l_2}
                 \frac{\log_{l_2}D_{B,j}}{\log_{l_2}l_1l_2}.
\end{eqnarray*}
\end{IEEEproof}

\medskip

\noindent{\bf Remark}: $E(A \otimes B)=E(B \otimes A)$ even though
$A \otimes B \neq B \otimes A$ in general.

\medskip

\setcounter{corollary}{13}

\begin{corollary}\label{corollary14}
The exponent of the polarizing matrix $A \otimes B$ is an
internally dividing point of $E(A)$ and $E(B)$. That is,
\begin{equation*}
E(A \otimes B)=\frac{\alpha}{1+\alpha}E(A)+\frac{1}{1+\alpha}E(B)
\end{equation*}
where $\alpha=\log_{l_2}l_1 \geq 0$.
\end{corollary}
\begin{IEEEproof}
By Theorem \ref{theorem13}, we have
\begin{equation*}
E(A \otimes
B)=\frac{E(A)}{1+\log_{l_1}l_2}+\frac{E(B)}{1+\log_{l_2}l_1}.
\end{equation*}
Without loss of generality, we may assume that $l_1 \leq l_2$. Let
$\alpha=\log_{l_2}{l_1}$. Then $0 \leq \alpha \leq 1$ and
\begin{eqnarray*}
E(A\otimes B)&=&\frac{1}{1+1/\alpha}E(A)+\frac{1}{1+\alpha}E(B)\\
&=&\frac{\alpha}{1+\alpha}E(A)+\frac{1}{1+\alpha}E(B).
\end{eqnarray*}
\end{IEEEproof}

\medskip

\begin{corollary}\label{corollary15}
Let $A_{1}$, $A_{2}$ be polarizing matrices of size $l_1 \times
l_1$ and let $B_{1}$, $B_{2}$ be polarizing matrices of size $l_2
\times l_2$. Assume that $E(A_1) \geq E(A_2)$ and $E(B_1) >
E(B_2)$, or $E(A_1) > E(A_2)$ and $E(B_1) \geq E(B_2)$. Then
\begin{equation*}
E(A_1 \otimes B_1) > E(A_2 \otimes B_2).
\end{equation*}
\end{corollary}

Corollary \ref{corollary15} tells us that a polarizing matrix with
higher exponent should be selected as a component matrix when we
construct a polarizing matrix with higher rate of polarization
from the Kronecker product.

\bigskip

Theorems \ref{theorem12} and \ref{theorem13} can be generalized to
a polarizing matrix $A=A_1 \otimes A_2 \otimes \cdots \otimes A_N$
of length $l=l_1 l_2 \cdots l_N$ where $A_{i}$ is an $l_i \times
l_i$ polarizing matrix for $i=1,2,\ldots,N$.

\setcounter{theorem}{15}

\medskip

\begin{theorem} \label{theorem16}
Let $k$ be an integer with $1 \leq k \leq l_1 l_2 \cdots l_N$.
Then the $k$th partial distance of the polarizing matrix $A_1
\otimes A_2 \otimes \cdots \otimes A_N$ is given by
\begin{equation*}
D_{A_1 \otimes A_2 \otimes \cdots \otimes A_N,k}=
D_{A_1,i_1}D_{A_2,i_2}\cdots D_{A_N,i_N}
\end{equation*}
where $k=(i_1-1)l_2l_3\cdots l_N+(i_2-1)l_3l_4\cdots
l_N+\cdots+(i_{N-1}-1)l_N+i_N$ with $1 \leq i_j \leq l_j$ for
$j=1,2,\ldots,N$.
\end{theorem}
\begin{IEEEproof}
Since the Kronecker product is associative, i.e., $A\otimes (B
\otimes C)=(A \otimes B) \otimes C$, the statement can be easily
derived in a recursive way.
\end{IEEEproof}

\medskip

\begin{theorem} \label{theorem17}
The exponent of the polarizing matrix $A_1 \otimes A_2 \otimes
\cdots \otimes A_N$ is given by
\begin{equation*}
E(A_1 \otimes A_2 \otimes \cdots \otimes A_N)= \sum_{i=1}^{N}
\frac{E(A_i)}{\log_{l_i}l_1 l_2 \cdots l_N}.
\end{equation*}
\end{theorem}
\begin{IEEEproof}
It is similar to the Proof of Theorem \ref{theorem13}.
\end{IEEEproof}

\medskip

\setcounter{corollary}{17}

\begin{corollary} \label{corollary 18}
Let $A$ be an $l \times l$ polarizing matrix. For any integer $N
\geq 1$, the exponent of the $N$th Kronecker power of $A$,
$A^{\otimes N}=A \otimes \cdots \otimes A$, is given by
\begin{equation*}
E\left(A^{\otimes N}\right)=E(A).
\end{equation*}
\end{corollary}



\section{Design Examples}

In order to illustrate the relationship between the exponent of a
polarizing matrix constructed from the Kronecker product and the
error rate of the corresponding polar code, some design examples
are presented in this section. The following matrices are employed
as a component matrix for larger polarizing matrices:
\[
G_{2}=\left[
\begin{array}{cc}
  1 & 0 \\
  1 & 1 \\
\end{array}
\right],~G_{3,L}=\left[
\begin{array}{ccc}
  1 & 0 & 0 \\
  1 & 0 & 1 \\
  1 & 1 & 1 \\
\end{array}
\right], \] \[G_{3,H}=\left[
\begin{array}{ccc}
  1 & 0 & 0 \\
  1 & 1 & 0 \\
  0 & 1 & 1 \\
\end{array}
\right]
\]
where $G_{2}$ is proposed by Ar{\i}kan \cite{Arikan1}, $G_{3,L}$
is introduced in \cite{Korada_IT} and $G_{3,H}$ is newly designed.
Using these matrices, we construct two polarizing matrices of size
$6 \times 6$ given by
\begin{eqnarray*}
    G_{6,L}=G_{2} \otimes G_{3,L},~~G_{6,H}=G_{2} \otimes G_{3,H}.
\end{eqnarray*}
The partial distances and the exponents of the above matrices are
given in Table \ref{partial_table}. Since $E(G_{3,H}) >
E(G_{3,L})$, we have $E(G_{6,H}) > E(G_{6,L})$ as shown in
Corollary \ref{corollary15}.

\begin{table}
 \caption{Partial distances and exponents of $G_{2}$ \cite{Arikan1},
     $G_{3,L}$\cite{Korada_IT}, $G_{3,H}$, $G_{6,L}$ and $G_{6,H}$.}
     \label{partial_table}
\begin{center}
\begin{tabular}{|c||c|c|}
    \hline
  ~~~Matrix~~~                     & ~~~Partial distances~~~     & ~~~Exponent~~~   \\
    \hline \hline
  $G_{2}$  \cite{Arikan1}          & $1,2$                       & $0.500$            \\
    \hline
  $G_{3,L} \cite{Korada_IT}$       & $1,1,3$                     & $0.333$           \\
   \hline
  $G_{3,H}$                        & $1,2,2$                     & $0.421$           \\
   \hline
  $G_{6,L}$                        & $1,1,3,2,2,6$               & $0.398$           \\
  \hline
  $G_{6,H}$                        & $1,2,2,2,4,4$               & $0.451$           \\
  \hline
\end{tabular}
\end{center}
\end{table}

\begin{table*}
 \caption{Exponents of $l \times l$ polarizing matrices
          constructed from the Kronecker product
          for $32 \leq l \leq 128$.}
     \label{exponent_large_polar}
\begin{center}
\begin{tabular}{|c|c|c||c|c|c||c|c|c|}
    \hline
       $l$             &  Matrix                         & Exponent
     & $l$             &  Matrix                         & Exponent
     & $l$             &  Matrix                         & Exponent \\
    \hline \hline
     $32$              &  $G_2 \otimes G_{{\rm S},16}$       & $0.5146$   &
     $56$              &  $G_2 \otimes G_{{\rm S},28}$       & $0.5121$   &
     $84$              &  $G_{3,H} \otimes G_{{\rm S},28}$   & $0.4914$      \\
    \hline
     $33$              &  $G_{3,H} \otimes G_{{\rm S},11}$   & $0.4492$     &
     $57$              &  $G_{3,H} \otimes G_{{\rm S},19}$   & $0.4694$     &
     $87$              &  $G_{3,H} \otimes G_{{\rm S},29}$   & $0.4935$        \\
    \hline
     $34$              &  $G_2 \otimes G_{{\rm S},17}$                   & $0.4934$      &
     $58$              &  $G_2 \otimes G_{{\rm S},29}$                   & $0.5142$      &
     $88$              &  $G_{2}^{\otimes 2} \otimes G_{{\rm S},22}$     & $0.4962$     \\
    \hline
     $36$              &  $G_2 \otimes G_{{\rm S},18}$       & $0.4917$      &
     $60$              &  $G_2 \otimes G_{{\rm S},30}$       & $0.5183$      &
     $90$              &  $G_{3,H} \otimes G_{{\rm S},30}$   & $0.4974$      \\
    \hline
     $38$              &  $G_2 \otimes G_{{\rm S},19}$                   & $0.4898$   &
     $62$              &  $G_2 \otimes G_{{\rm S},31}$                   & $0.5220$   &
     $92$              &  $G_{2}^{\otimes 2} \otimes G_{{\rm S},23}$     & $0.5005$      \\
    \hline
     $39$              &  $G_{3,H} \otimes G_{{\rm S},13}$   & $0.4635$  &
     $63$              &  $G_{3,H} \otimes G_{{\rm S},21}$   & $0.4695$  &
     $93$              &  $G_{3,H} \otimes G_{{\rm S},31}$   & $0.5009$      \\
    \hline
     $40$              &  $G_2 \otimes G_{{\rm S},20}$                 & $0.4972$    &
     $64$              &  $G_{2}^{\otimes 2} \otimes G_{{\rm S},16}$   & $0.5122$    &
     $96$              &  $G_{2}^{\otimes 2} \otimes G_{{\rm S},24}$   & $0.5031$      \\
    \hline
     $42$              &  $G_2 \otimes G_{{\rm S},21}$                  & $0.4895$   &
     $66$              &  $G_{3,H} \otimes G_{{\rm S},22}$              & $0.4752$   &
     $100$             &  $G_{2}^{\otimes 2} \otimes G_{{\rm S},25}$    & $0.5003$       \\
    \hline
     $44$              &  $G_2 \otimes G_{{\rm S},22}$                 & $0.4955$    &
     $68$              &  $G_{2}^{\otimes 2} \otimes G_{{\rm S},17}$   & $0.4945$    &
     $104$             &  $G_{2}^{\otimes 2} \otimes G_{{\rm S},26}$   & $0.5033$      \\
    \hline
     $45$              &  $G_{3,H} \otimes G_{{\rm S},15}$             & $0.4756$    &
     $69$              &  $G_{3,H} \otimes G_{{\rm S},23}$             & $0.4800$    &
     $108$             &  $G_{2}^{\otimes 2} \otimes G_{{\rm S},27}$   & $0.5059$      \\
    \hline
     $46$              &  $G_2 \otimes G_{{\rm S},23}$                 & $0.5006$    &
     $72$              &  $G_{2}^{\otimes 2} \otimes G_{{\rm S},18}$   & $0.4930$    &
     $112$             &  $G_{2}^{\otimes 2} \otimes G_{{\rm S},28}$   & $0.5103$      \\
    \hline
     $48$              &  $G_2 \otimes G_{{\rm S},24}$                 & $0.5037$    &
     $75$              &  $G_{3,H} \otimes G_{{\rm S},25}$             & $0.4802$    &
     $116$             &  $G_{2}^{\otimes 2} \otimes G_{{\rm S},29}$   & $0.5121$       \\
    \hline
     $50$              &  $G_2 \otimes G_{{\rm S},25}$                 & $0.5003$    &
     $76$              &  $G_{2}^{\otimes 2} \otimes G_{{\rm S},19}$   & $0.4914$    &
     $120$             &  $G_{2}^{\otimes 2} \otimes G_{{\rm S},30}$   & $0.5157$     \\
    \hline
     $51$              &  $G_{3,H} \otimes G_{{\rm S},17}$                 & $0.4720$   &
     $78$              &  $G_{3,H} \otimes G_{{\rm S},26}$                 & $0.4836$   &
     $124$             &  $G_{2}^{\otimes 2}     \otimes G_{{\rm S},31}$   & $0.5188$      \\
    \hline
     $52$              &  $G_2 \otimes G_{{\rm S},26}$                     & $0.5039$   &
     $80$              &  $G_{2}^{\otimes 2} \otimes G_{{\rm S},20}$       & $0.4977$   &
     $126$             &  $G_{2} \otimes G_{3,H} \otimes G_{{\rm S},21}$   & $0.4737$       \\
    \hline
     $54$              &  $G_2 \otimes G_{{\rm S},27}$                     & $0.5069$    &
     $81$              &  $G_{3,H} \otimes G_{{\rm S},27}$                 & $0.4865$    &
     $128$             &  $G_{2}^{\otimes 3}     \otimes G_{{\rm S},16}$   & $0.5104$     \\
    \hline
\end{tabular}
\end{center}
\end{table*}

We designed four half-rate polar codes whose generator matrices
are $G_{6,L}^{\otimes 4}$, $G_{6,H}^{\otimes 4}$,
$G_{6,L}^{\otimes 5}$, $G_{6,H}^{\otimes 5}$, respectively, and
whose frozen bits are optimized to the binary erasure channel with
erasure rate $1/2$.{\footnote[1]{We employ Ar{\i}kan's heuristic
method \cite{Arikan3} to find the frozen bits.}} It is assumed
that the coded bits are modulated to binary phase-shift keying
(BPSK) symbols and then transmitted over an additive white
Gaussian noise (AWGN) channel. Fig. \ref{fig1_BLER} shows the
block error rates of these polar codes under SC decoding, where
$E_{b}$ is the received signal energy per information bit and
$N_{0}$ is the one-sided power spectral density of the AWGN. The
polar codes with $G_{6,H}$ as a component polarizing matrix have
much lower error rates than those with $G_{6,L}$ in the high
signal-to-noise power ratio (SNR) region. This result shows that
when a polarizing matrix is constructed from the Kronecker
product, it is required to select a polarizing matrix with high
exponent as a component matrix.

\begin{figure}[t]
 \center
   \includegraphics[width=0.5\textwidth]{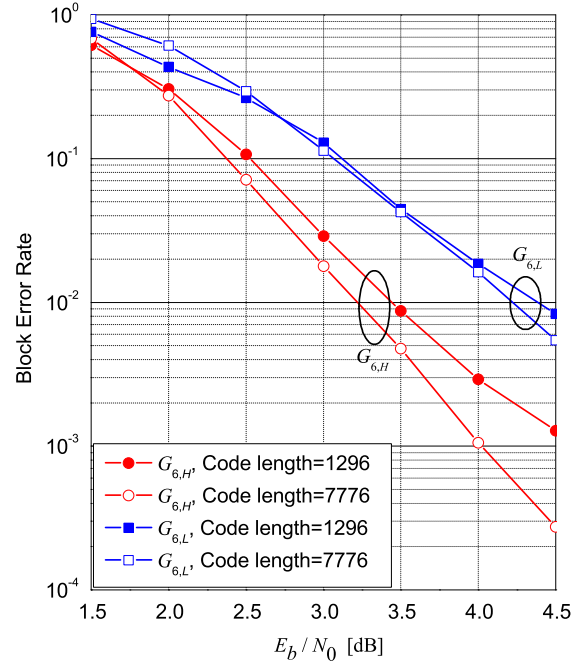}
   \caption{Block error rates of half-rate polar codes whose lengths
            are 1296 and 7776 bits over
            an AWGN channel.}
   \label{fig1_BLER}
\end{figure}


Korada {\it et al.} \cite{Korada_IT} constructed new polarizing
matrices of size $m \times m$ for $m \leq 31$ obtained by
shortening a BCH code of length $31$. For our reference, we denote
such an $m \times m$ matrix by $G_{{\rm S},m}$. The exponent of
$G_{{\rm S},m}$ provides a lower bound on the maximum exponent for
polarizing matrices of size $m \times m$, defined as
\begin{equation*}
    E_{m} \triangleq \max_{G \in \{0,1\}^{m \times m}} E(G),
\end{equation*}
in a constructive way. Note that polarizing matrices with $m
> 31$ may be constructed from the method proposed in
\cite{Korada_IT}. However, it is a very difficult problem to
calculate their exponents, since a search space for computing
their partial distances becomes significantly large. The
difficulty may be overcome by employing the Kronecker product. As
an example, for $32 \leq l \leq 128$,{\footnote[2]{For a simple
example, the size $l$ is restricted to $32 \leq l \leq 128$.
Polarizing matrices of size $l \times l$ for $l \geq 129$ can be
constructed in a similar approach.}} the exponents of $l \times l$
polarizing matrices of the form $G_{2}^{\otimes n_1} \otimes
G_{3,H}^{\otimes n_2} \otimes G_{{\rm S},m}$ are easily calculated
by Theorem \ref{theorem13} and are presented in Table
\ref{exponent_large_polar}. Note that these exponents may become a
good lower bound on $E_l$ for $32 \leq l \leq 128$.

\bigskip

\section{Conclusions}

We derived the partial distances and the exponent of a polarizing
matrix constructed from the Kronecker product. Our results can be
employed in the design of a polarizing matrix with high exponent
when it is constructed from the Kronecker product. It is expected
that our approach can be generalized to the calculation of the
partial distances and the exponent of a nonbiary polar code.


\enlargethispage{-2.6in}

\end{document}